\documentstyle[11pt,moriond,epsfig]{article}

\def\Journal#1#2#3#4{{#1} {\bf #2}, #3 (#4)}

\def\PRL{\em Phys. Rev. Lett.}

\def\be{\begin{equation}}
\def\ee{\end{equation}}
\def\bea{\begin{eqnarray}}
\def\eea{\end{eqnarray}}

\begin{document}
\vspace*{4cm}
\title{A Study of B meson Decays to Charm at Belle}

\author{ C.H. Wang (Belle Collaboration)}

\address{National Lien-Ho Institude of Technology, 
Miao-Li, Taiwan, R.O.C.}

\maketitle\abstracts{With ~10fb$^{-1}$ data from Belle, the results 
of b to c and u decays have been represented here. Several decay 
modes have been studied, which include charmonium, 
semileptonic decay and Cabibbo-suppressed decays mode.
 The preliminary results and 
the measurements of KM matrix elements will also be presented here.}

\section{Introduction}

B decays provide a very rich ground for the test of the 
Standard Model(SM). The Kobayashi-Maskawa(KM) matrix is incorporated
in weak decays sector of the Standard Model to explain the 
quark mixing and CP violation. 
Hence, the experimental measurements of all the KM matrix elements 
will either complete our understanding on the weak interaction or 
lead to the new physics beyond the Standard Model.

The charmonium mesons from B decays 
play an important role in the
studies of CP violation phemomena. 
Their production mechanism also 
provides the test ground for low energy QCD. 
The Cabibbo-suppressed modes $B \to D K$ can provide a theoretically 
clean method to determine $\phi_3$(also known as $\gamma$). Ratio of 
the Cabiboo-suppressed $B \to D^{(*)} K$ to the Cabiboo-allowed 
$B \to D^{(*)} \pi$ will give us 
information on Cabiboo mixing angle, pion and kaon decay constants
assuming factorization and SU(3) symmetry. 
The measurements of B semileptonic decays branching ratio 
will give us the measurements of $V_{cb}$ and $V_{ub}$, 
another KM matrix elements.

In this paper, the KEKB storage ring, Belle detector and its performance
will not be described here. The complete description can be found in 
reference elsewhere\cite{HARD}.
This paper will be organized as the follows. The brief analysis 
techniques will 
be described in the section 2. Following that section will be the 
preliminary results of selected decay modes, where the more detailed 
analysis information can be 
found in http://belle.kek.jp/belle/publications.html.

\section{Analysis}
\subsection{General Features}

The results are based on the 10.5 $fb^{-1}$ data collected on the 
$\Upsilon(4S)$ energy(resonance) and 0.6 $fb^{-1}$ data on the
energy 60 $MeV$ lower than the $\Upsilon(4S)$ resonance energy(off-resonance).

For exclusive decays, B candidates are identified through the beam 
constrained mass $M_B = \sqrt{E_{beam}-{p_B}^*}$ 
and $\Delta E = E_{beam} - E_B $, where 
$E_B$, ${p_B}^*$ are the B energy and momentum 
in CM frame and $E_{beam}=E_{CMS}/2$.
For the two body decay modes, the daughter particles' momentum 
will generally fall between 2 and 3 GeV in CM frame 
which is very rare in general b to c decay. 
Therefore, the major dominant background will come from 
continuum events which is 
from $e\bar{e} \to \gamma^* \to q\bar{q}$ process. 
Due to the production diagram, the continuum event has a two-jet 
structure and is very different from the $\Upsilon(4S) \to B\bar{B}$ 
process, where B decay almost at rest (spherical shape) 
in $\Upsilon(4S)$ rest(CM) frame. The event shape
variables are then used to suppress continuum background. 

For B inclusive and B semileptonic decays, the continuum background 
is subtracted through the results using off-resonance data with 
proper luminosity and kinematic scaling. 

\subsection{Charmonium Production}

For direct charmonium production, the momentum of that charmonium 
in the CM frame is required to be greater than 2 $GeV/c$, 
which kinematically eliminate the charmonium from B decays. 
The charged track multiplicity is greater than 4. 
The charmonium modes we reconstructed are
$J/\psi \to l^+ l^-$, $\psi(2S) \to l^+ l^-$ ($l=e$ or $\mu$),
$\psi(2S) \to J/\psi \pi^+ \pi^-$ and $\chi_c \to J/\psi \gamma$.
For $J/\psi$ mode, with $530\pm 43$ from the resonance data and
$30 \pm 7$ from the off-resonance data, we got $17.5 \pm 127$ after
off-resonance data subtraction. The results show no 
direct $J\psi$ production from $\Upsilon(4S)$ with the limit
$Br(\Upsilon(4S) \to J/\psi X)<4.1\times 10^{-4}$.
This is incompatible with the CLEO 
measurement:$(2.2 \pm 0.6\pm0.4)\times 10^{-3}$.
All the observed ones are consistent with 
from $e\bar{e} \to \gamma^* $ process. We obtained 
$\sigma(e^+e^- \to q \bar{q} \to J/\psi X) = (1.02 \pm 0.08 \pm 0.12)$ pb.
The ratio R of $J/\psi$, which is 
$\sigma(e^+e^- \to q \bar{q} \to J/\psi X)/\sigma(e^+e^- \to \mu^+\mu^-)$, 
is $(1.32 \pm 0.14 \pm 0.15)\times 10^{-3}$. 
The results for direct $\psi(2S) X$
is $\sigma(e^+e^- \to q \bar{q} \to \psi(2S) X) = (0.54 \pm 0.12 )$ pb.
Both the production 
cross-section of $J/\psi$ and $\psi(2S)$ agree well 
with NRQCD calculation\cite{NRQCD}. 

For inclusive charmonium from B decays, the charmonium momentum in CM frame 
is required to be less than 1.7 $GeV/c$. $\chi_c$ is identified through
mass difference $m({l^+l^- \gamma}) - m(l^+l^-)$. Charge multiplicity 
and off-resonance data subtraction are also applied. Results are 
:$Br(B \to J/\psi X)=(0.25 \pm 0.04 \pm 0.03) \times 10^{-2}$ ,
$Br(B \to \psi^\prime(l^+l^-) X)=(0.25 \pm 0.04 \pm 0.03) \times 10^{-2}$,
$Br(B \to \psi^\prime(\psi\pi\pi) X)=(0.31 \pm 0.04 \pm 0.04)\times 10^{-2}$,
$Br(B \to \chi_{c1} X)=(0.39 \pm 0.03 \pm 0.05)\times 10^{-4}$,
$Br(B \to \chi_{c2} X)=(0.18 \pm 0.05 \pm 0.02)\times 10^{-4}$.

\subsection{$B \to J/\psi K_1(1270)$}

\begin{figure}
\epsfig{figure=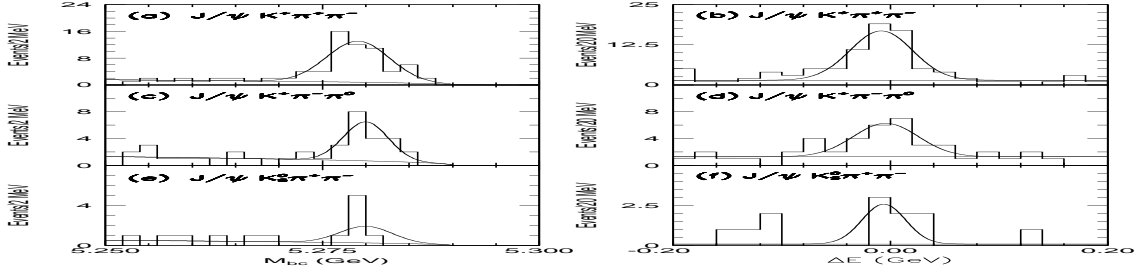,height=3.5cm,width=15.0 cm}
\caption{$M_B$ and $\Delta E$ for $J/\psi K_1(1270)$ after $K_1$ mass cuts.}
\label{pol}
\end{figure}

In this decay modes, $K_1(1270)$\cite{psik1} is reconstructed through 
modes: $K^+ \pi^+ \pi^-$, $K^+ \pi^- \pi^0$ and $K_s \pi^+ \pi^-$.
The feed-down from $\psi(2s) \to J/\psi \pi \pi$ are also excluded here.
The yields are extracted through simultaneous fits on $M_B$ 
and $\Delta E$ of B candidate while 
cutting $K \pi \pi$ in the $K_1(1270)$ mass range.
The results are shown in Fig.~\ref{pol} and the fitted numbers are 
$53.4 \pm 9.1$ events for $J/\psi K^+ \pi^- \pi^-$, $19.3 \pm 5.1$
events for $J/\psi K^+ \pi^- \pi^0$ and $6.2 \pm 2.6$
events for $J/\psi K^0 \pi^+ \pi^-$. The branching ratio are:
$Br(B \to J/\psi K_1^0(1270))\over Br(B \to J/\psi K^+)$ 
$=1.30 \pm 0.34 \pm 0.28$ , 
$Br(B \to J/\psi K_1^+(1270))\over Br(B \to J/\psi K^+)$ 
$=1.80 \pm 0.34 \pm 0.34$  

\subsection{Cabiboo Suppressed $B \to D^{(*)} K$ decays}

\begin{figure}
\epsfig{figure=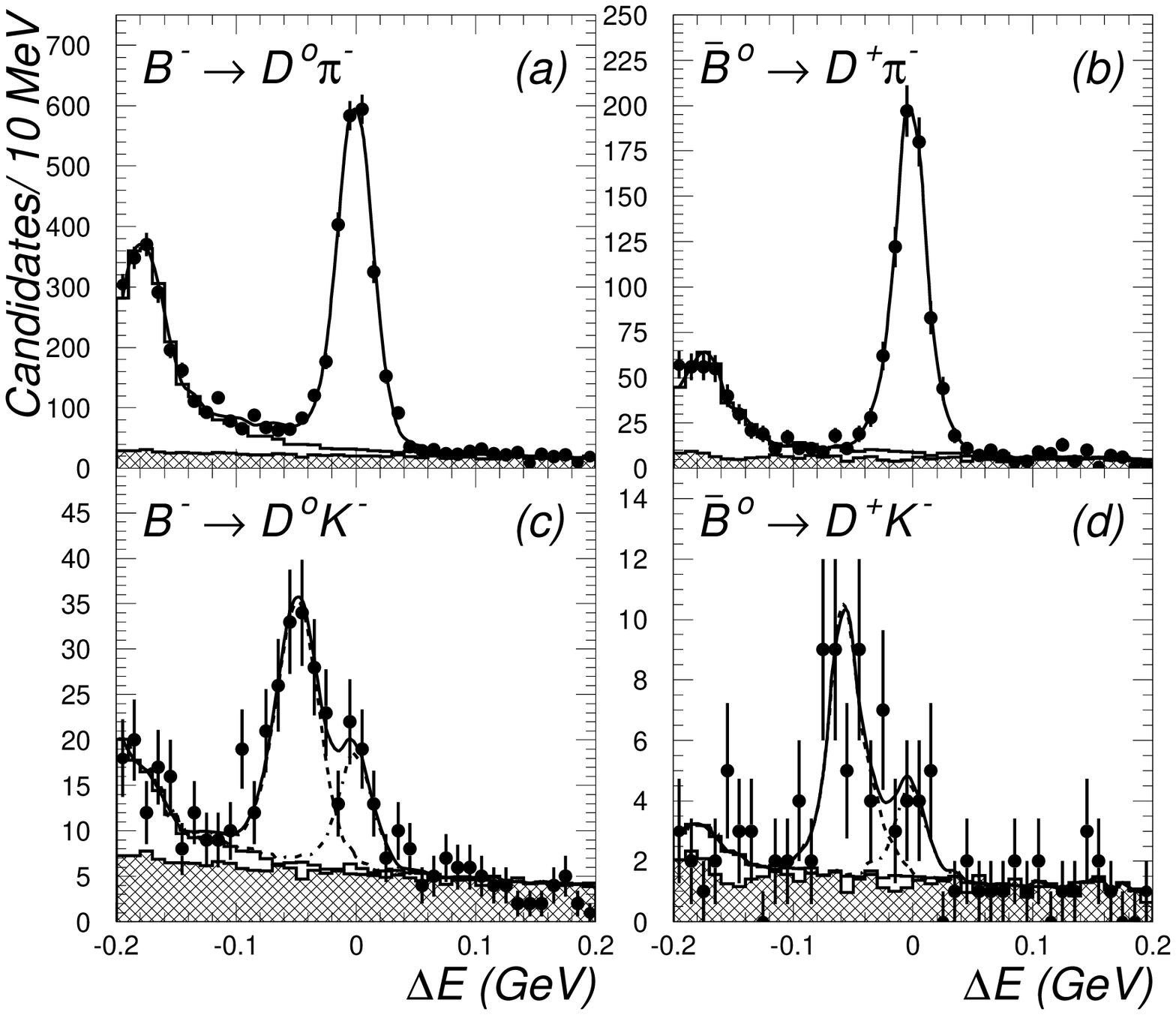,height=3.8cm,width=7.5 cm}
\epsfig{figure=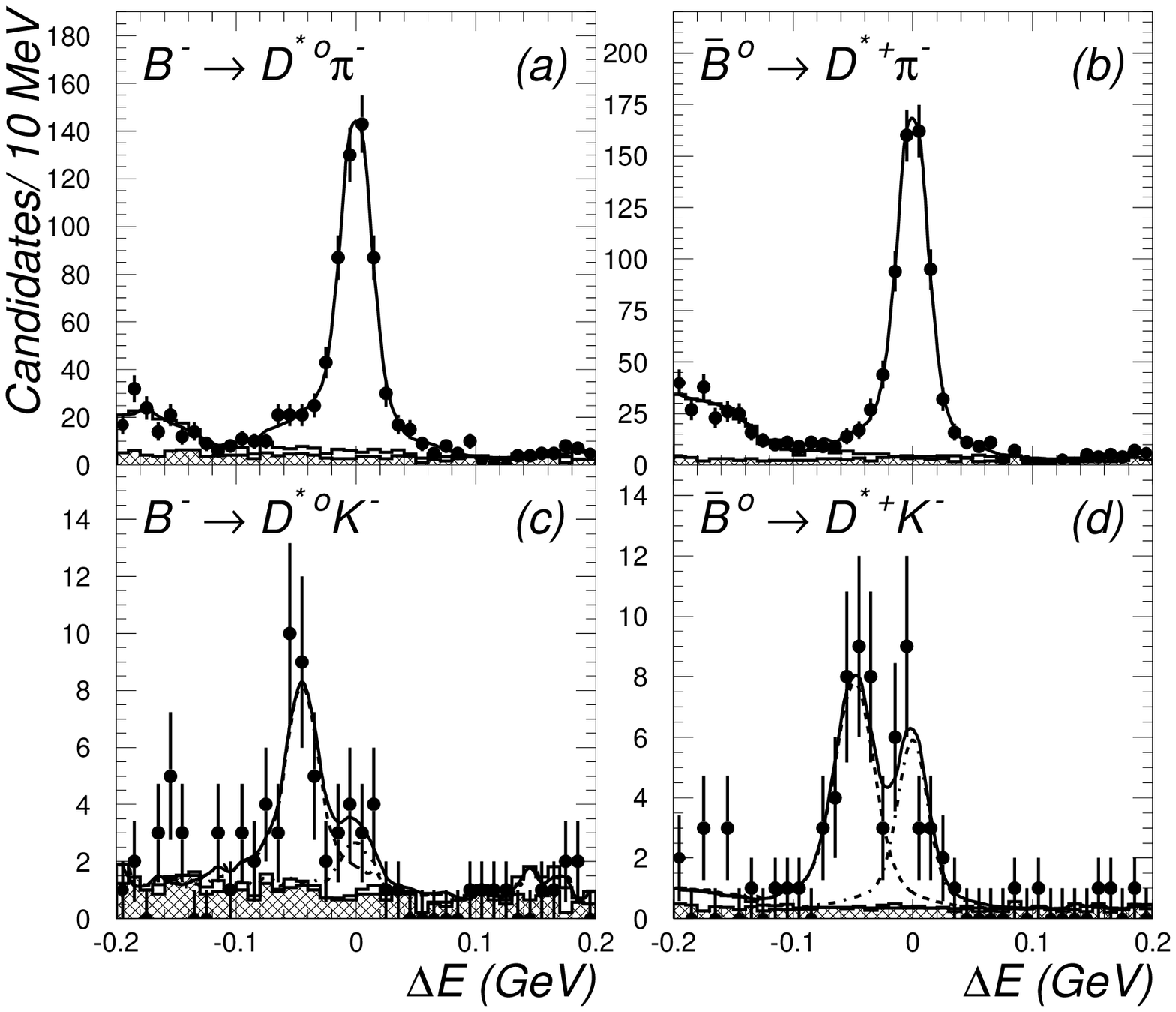,height=3.8cm,width=7.5 cm}
\caption{$\Delta E$ distribution for $D K^+$ and $D^* K^+$.}
\label{dks}
\end{figure}

Due to the good particle identification system in Belle, 
we can easily reduce the Cabiboo favored $B \to D^{(*)} \pi$ feed-down 
and extract the $B \to D^{(*)} K$. The $D^{*+}$ and $D^{*0}$ candidate 
are reconstructed through $D^{*0} \to D^0 \pi^0$, $D^{*+} \to D^0 \pi^+$ 
and $D^+ \pi^0$ with $D^0 \to K^- \pi^+ , K^- \pi^+ \pi^0 ,
K^- 3(\pi^\pm)$ and $D^+ \to K^- \pi^+ \pi^+ , K_s \pi^+ , 
K_s \pi^+ \pi^0, K_s \pi^+ \pi^+ \pi^-$.
The $K^{*+}$ is reconstructed through $K_s \pi^+$.
The invariant mass of D is required to be within 4$\sigma$ of 
the known D mass. Vertex cuts will also be applied to get a clean D sample. 
The $\Delta E$ distribution is shown in Fig.~\ref{dks}.
Results are $Br(B^- \to D^0 K^-)\over Br(B^- \to D^0 \pi^-)$ 
$=0.079 \pm 0.009 \pm 0.006$,  
$Br(\bar{B^0} \to D^+ K^-)\over Br(\bar{B^0} \to D^+ \pi^-)$ 
$=0.068 \pm 0.015 \pm 0.007$, 
$Br(B^- \to D^{*0} K^-)\over Br(B^- \to D^{*0} \pi^-)$ 
$=0.078 \pm 0.019 \pm 0.009$ and
$Br(\bar{B^0} \to D^{*+} K^-)\over Br(\bar{B^0} \to D^{*+} \pi^-)$ 
$=0.074 \pm 0.015 \pm 0.006$.

\subsection{$|V_{cb}|$ measurements}

\begin{figure}
\epsfig{figure=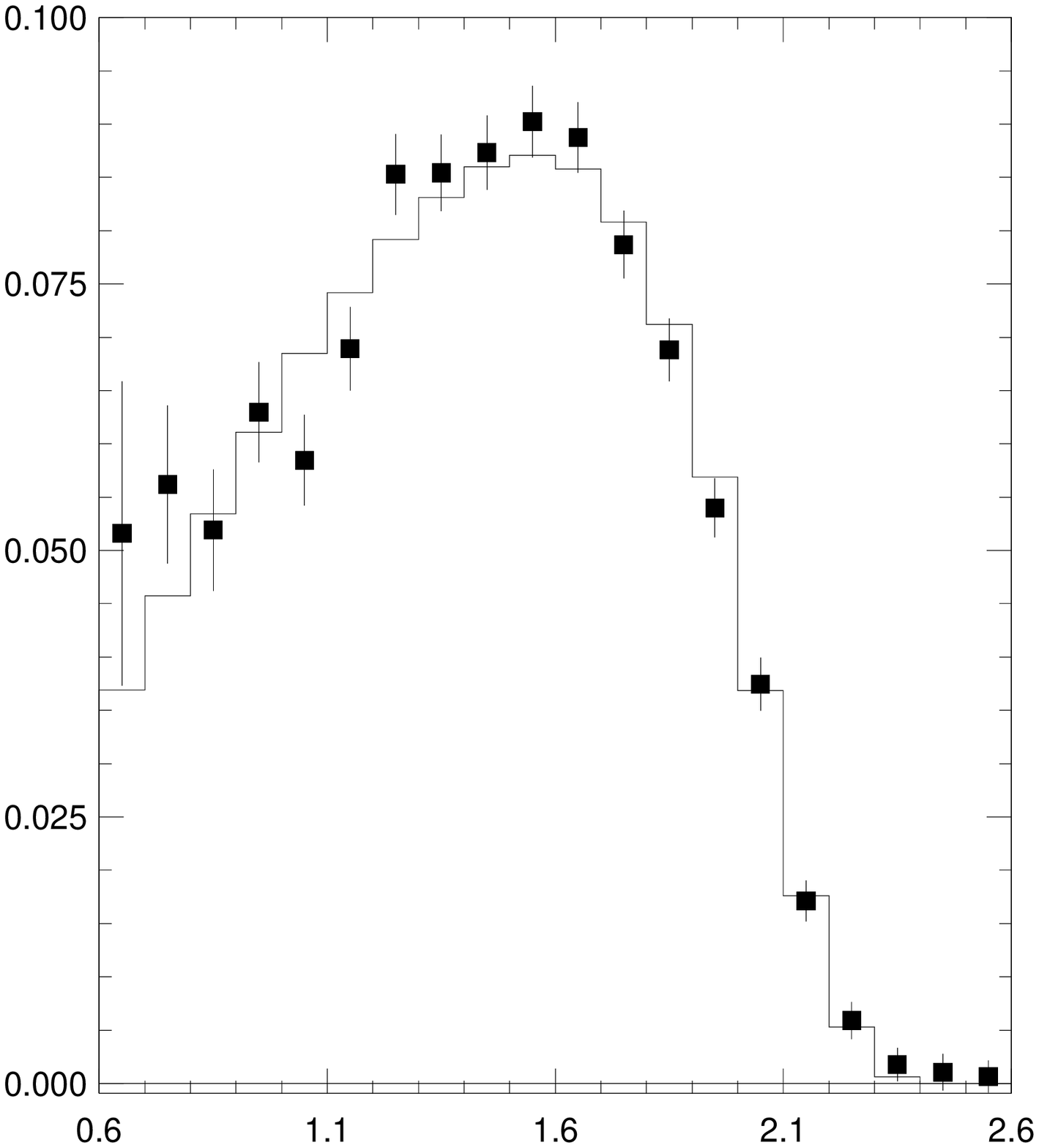,height=3.8cm,width=7.5 cm}
\epsfig{figure=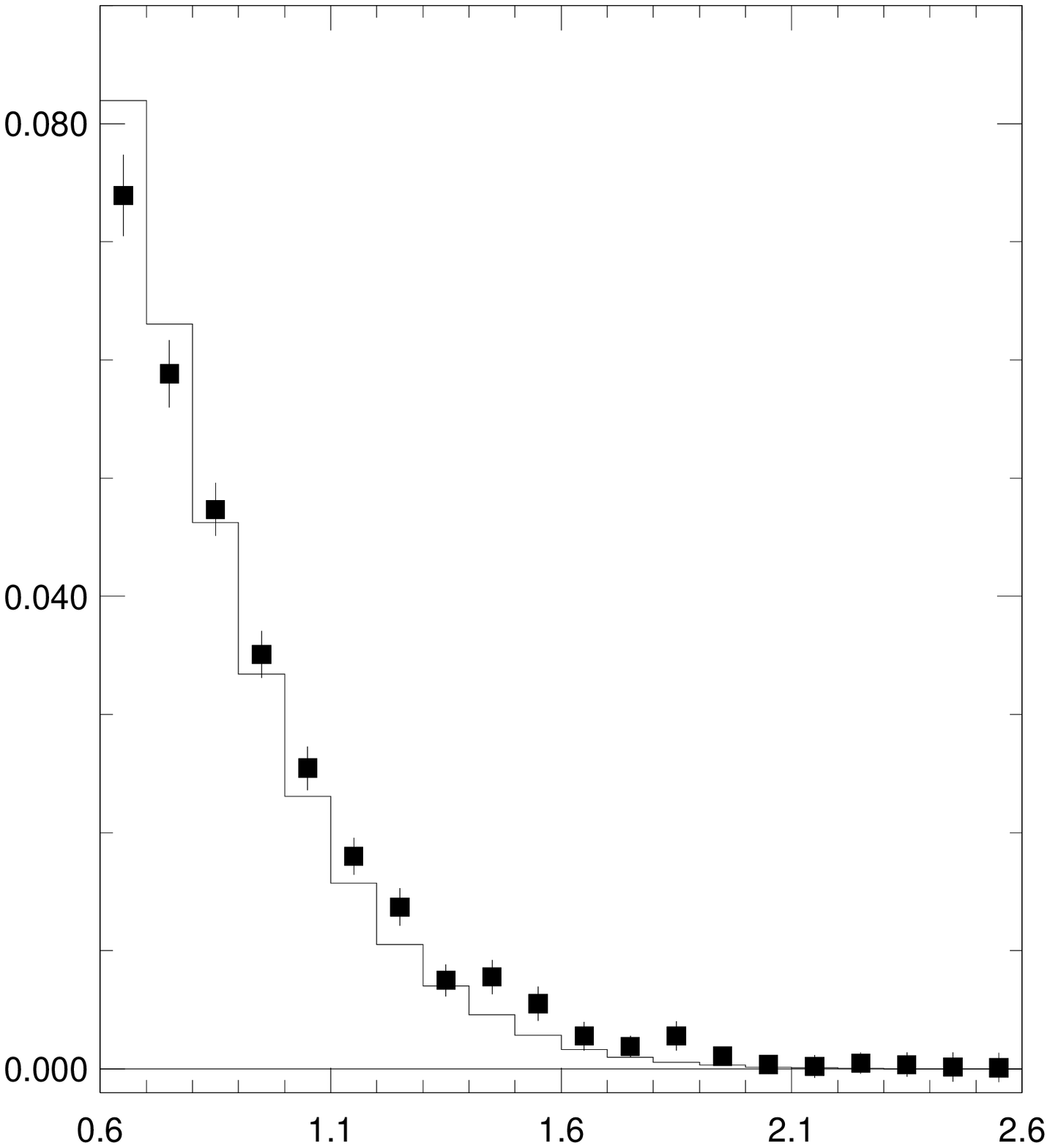,height=3.8cm,width=7.5 cm}
\caption{Primary and secondary electron spectrum.}
\label{tag}
\end{figure}

Data used in these analysis ranges from 5.1 $pb^{-1}$ to 5.8 $pb^{-1}$. 
There are basically two methods for this measurement. One is to measure
the lepton spectrum from dilepton events by tagging one lepton 
within $1.5 < P^*_l < 2.2$ $GeV/c$. The primary and the
secondary lepton spectrum are then fitted simultanously (Fig.~\ref{tag}.
The branching fraction we measure is 
$Br(B\to X e \bar{\nu_e})= (11.05 \pm 0.15 \pm 0.46)$\%. The $|V_{cb}|$ 
is extracted through the relation 
$|V_{cb}|^2 = {Br(B\to x e \bar{\nu_e} \over (\tau_B\gamma_c)}$, which give
$|V_{cb}|=(4.15\pm 0.09 \pm 0.26)\times 10^{-2}$ with $\gamma_c=40 \pm 8$
from ACCMM model and $|V_{cb}|=(4.05\pm 0.09 \pm 0.24)\times 10^{-2}$ 
with $\gamma_c=42 \pm 8$ from ISGW$**$ model. 

\begin{figure}
\epsfig{figure=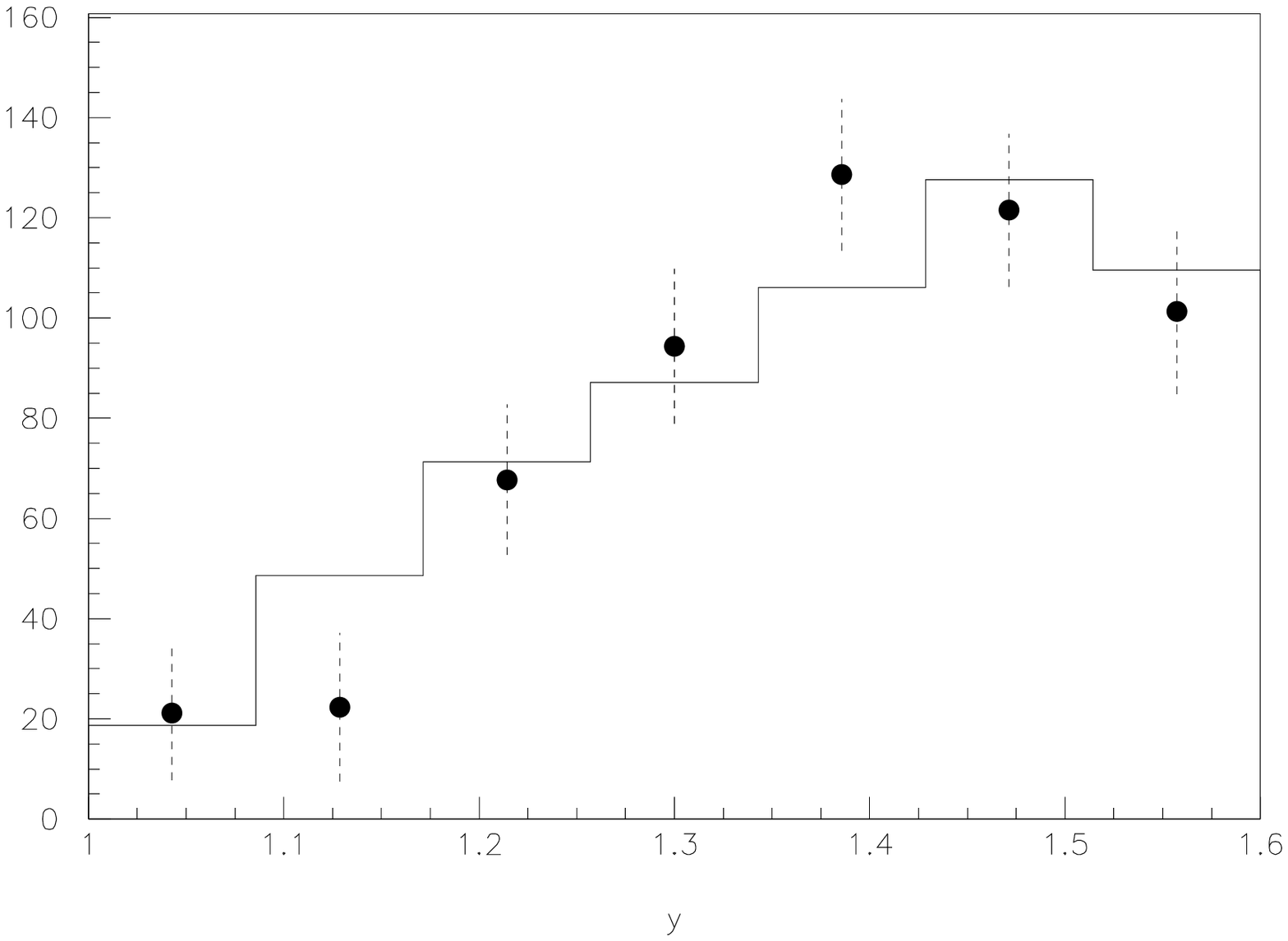,height=3.5cm,width=7.5 cm}
\epsfig{figure=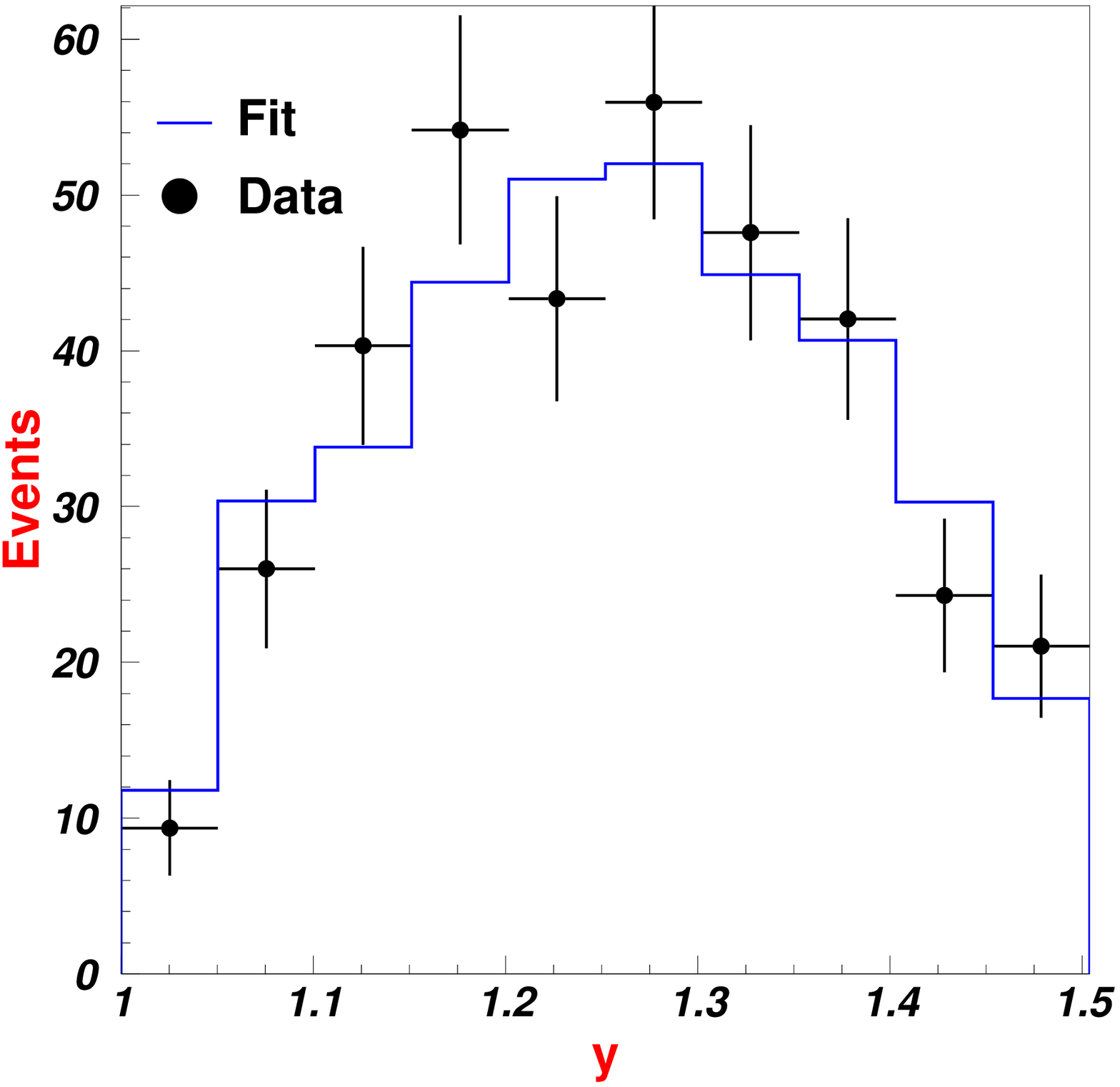,height=3.3cm,width=7.0 cm}
\caption{The yields as a function of y for $D$ and $D^*$ respectively.}
\label{vcb}
\end{figure}

Another method is based on the Heavy Quark Effect Theory (HQET)\cite{HQET} which
provide a accurate estimation of $|V_{cb}$ through form factor 
at the zero recoil energy limit. The form factor can be approximated
as $F_{D^{(*)}}(y) = F_D^{(*)}(1)[1-\rho^2_{D^{(*)}}(y-1)$. We fit $\rho^2$ and 
$|V_{cb}|F_{D^{(*)}}(1)$ simultanously. Assuming a linear form factor,
we obtain $|V_{cb}|F_{D^{(*)}}(1)=(3.57 \pm 0.11 \pm 0.13)\times 10^{-2}$
, $\rho^2_{D^{*}} = 0.93 \pm 0.02 \pm 0.02$ for $B \to D^* l \nu$. 
Using $F(1)_{D^*} = 0.913 \pm 0.042$, we have $|V_{cb}| = (3.91 \pm 0.12(stat)
\pm 0.15(sys)\pm 0.16(theory))\times 10^{-2}$. For $D l \nu$ mode,
we use $F(1)_D = 1.0 \pm 0.07$ and obtain 
$|V_{cb}|F_{D(1)}=(4.42 \pm 0.48 \pm 0.35)\times 10^{-2}$
, $\rho^2_{D^{*}} = 0.89 \pm 0.14 \pm 0.06$ 
and $|V_{cb}| = (4.42 \pm 0.48(stat)
\pm 0.35(sys)\pm 0.30(theory))\times 10^{-2}$. Figure~\ref{vcb} show the fits
results.

\section{Conclusion}

With $~$ 11 fb$^{-1}$ data, various decay modes have been study which 
include charmonium mesons, Carbiboo-suppress $D^{(*)} K$ and 
$|V_{cb}|$ measurement. Results are
$Br(\Upsilon(4S) \to J/\psi X)<4.1\times 10^{-4}$,
$\sigma(e^+e^- \to q \bar{q} \to J/\psi X) = (1.02 \pm 0.08 \pm 0.12)$ pb,
$\sigma(e^+e^- \to q \bar{q} \to \psi(2S) X) = (0.54 \pm 0.12 )$ pb,
$Br(B \to J/\psi K_1^0(1270))\over Br(B \to J/\psi K^+)$ 
$=1.30 \pm 0.34 \pm 0.28$ and
$Br(B \to J/\psi K_1^+(1270))\over Br(B \to J/\psi K^+)$ 
$=1.80 \pm 0.34 \pm 0.34$ for charmonium meson.
From dilepton analysis, we have
$Br(B\to X e \bar{\nu_e})= (11.05 \pm 0.15 \pm 0.46)$\% and
obtain $|V_{cb}|F_{D^{(*)}}(1)=(3.57 \pm 0.11 \pm 0.13)\times 10^{-2}$
,$|V_{cb}|F_{D(1)}=(4.42 \pm 0.48 \pm 0.35)\times 10^{-2}$ from
$D^* l \nu$ and $D l \nu$ analysis respectively.
    
\section*{Acknowledgments}

We wish to thank the KEK accelerator group for excellent operations.
We acknowledge support from the Ministry of Education of Japan;
the Australian Research Council and Department of 
Industry, Science and Resources;
the Department of Science and Technology of India;
the BK21 program of the Ministry of Education of Korea;
the Polish State Committee for Scientific Research
under contract No.2P03B 17017;
the Ministry of Science and Technology of Russian Federation;
the National Science Council and the Ministry of Education of Taiwan;
the Japan-Taiwan Cooperative Program of the Interchange Association;
and the U.S. Department of Energy.

\end{document}